\pdfoutput=1

\documentclass[12pt,a4paper]{article}

\usepackage{ifthen} 
\newboolean{pdflatex}
\setboolean{pdflatex}{true} 

\newboolean{articletitles}
\setboolean{articletitles}{true} 

\newboolean{uprightparticles}
\setboolean{uprightparticles}{false} 

\def\paperauthors{} 
\def\paperasciititle{Search for Majorana neutrinos in W+->mu+mu+-jet decays} 
\def\papertitle{Search for heavy neutral leptons\\ in $W^+\to\mu^{+}\mu^{\pm}\text{\,jet}$ decays} 
\def\paperkeywords{{High Energy Physics}, {LHCb}} 
\def\papercopyright{\the\year\:  CERN for the benefit of the LHCb collaboration} 
\def\paperlicence{CC BY 4.0 licence}
\def\paperlicenceurl{https://creativecommons.org/licenses/by/4.0/}

\def\madgraph {\mbox{\textsc{MadGraph}}\xspace}

\usepackage{siunitx}
\DeclareSIUnit\eVperc{\eV\per\clight}
\DeclareSIUnit\clight{\text{\ensuremath{c}}}
\DeclareSIUnit\gevcc{\giga\eV\per\clight\squared}
\DeclareSIUnit\tev{\tera\eV}
\DeclareSIUnit\invfb{\per\femto\barn}
\DeclareSIUnit{\mum}{\micro\meter}

\usepackage{subfig}

\def\eq {Eq.~}
\def\tab {Table~}
\def\fig {Fig.~}
\def\sect {Section~}

\def\muonW      {{\ensuremath{\Pmu_{\W}}}\xspace}
\def\muonN      {{\ensuremath{\Pmu_{N}}}\xspace}
\def\jet      {{\ensuremath{\text{\,jet}}}\xspace}

\usepackage{xcolor}

\usepackage{booktabs}
\usepackage{collcell}

\usepackage[top=1in, bottom=1.25in, left=1in, right=1in]{geometry}

\usepackage{microtype}
\usepackage{lineno}  
\usepackage{xspace} 
\usepackage{caption} 

\usepackage{graphicx}  
\usepackage{color}
\usepackage{colortbl}
\graphicspath{{./figs/}} 

\usepackage{amsmath} 
\usepackage{amssymb}
\usepackage{amsfonts}
\usepackage{upgreek} 

\newcommand*\patchAmsMathEnvironmentForLineno[1]{%
\expandafter\let\csname old#1\expandafter\endcsname\csname #1\endcsname
\expandafter\let\csname oldend#1\expandafter\endcsname\csname
end#1\endcsname
 \renewenvironment{#1}%
   {\linenomath\csname old#1\endcsname}%
   {\csname oldend#1\endcsname\endlinenomath}%
}
\newcommand*\patchBothAmsMathEnvironmentsForLineno[1]{%
  \patchAmsMathEnvironmentForLineno{#1}%
  \patchAmsMathEnvironmentForLineno{#1*}%
}
\AtBeginDocument{%
\patchBothAmsMathEnvironmentsForLineno{equation}%
\patchBothAmsMathEnvironmentsForLineno{align}%
\patchBothAmsMathEnvironmentsForLineno{flalign}%
\patchBothAmsMathEnvironmentsForLineno{alignat}%
\patchBothAmsMathEnvironmentsForLineno{gather}%
\patchBothAmsMathEnvironmentsForLineno{multline}%
\patchBothAmsMathEnvironmentsForLineno{eqnarray}%
}

\usepackage{hyperxmp}

\usepackage[pdftex,
            pdfauthor={\paperauthors},
            pdftitle={\paperasciititle},
            pdfkeywords={\paperkeywords},
            pdfcopyright={Copyright (C) \papercopyright},
            pdflicenseurl={\paperlicenceurl}]{hyperref}

\usepackage[colorinlistoftodos,textsize=scriptsize]{todonotes}

\usepackage[bottom,flushmargin,hang,multiple]{footmisc}

\usepackage[all]{hypcap} 

\usepackage{xspace} 
\usepackage{upgreek}

\def\lhcb   {\mbox{LHCb}\xspace}

\def\MagUp {\mbox{\em Mag\kern -0.05em Up}\xspace}

\ifthenelse{\boolean{uprightparticles}}%
{

 \def\Pmu         {\ensuremath{\upmu}\xspace}                 
 \def\Pnu         {\ensuremath{\upnu}\xspace}

 \def\Ptau        {\ensuremath{\uptau}\xspace}

 \def\Ppsi        {\ensuremath{\uppsi}\xspace}

 \def\PDelta      {\ensuremath{\Delta}\xspace}                 
 \def\PXi         {\ensuremath{\Xi}\xspace}                 
 \def\PLambda     {\ensuremath{\Lambda}\xspace}                 
 \def\PSigma      {\ensuremath{\Sigma}\xspace}                 
 \def\POmega      {\ensuremath{\Omega}\xspace}                 
 \def\PUpsilon    {\ensuremath{\Upsilon}\xspace}

 \def\PB      {\ensuremath{\mathrm{B}}\xspace}                 
                  
 \def\PD      {\ensuremath{\mathrm{D}}\xspace}

 \def\PJ      {\ensuremath{\mathrm{J}}\xspace}                 
 \def\PK      {\ensuremath{\mathrm{K}}\xspace}

 \def\PW      {\ensuremath{\mathrm{W}}\xspace}

 \def\PZ      {\ensuremath{\mathrm{Z}}\xspace}                 
                  
 \def\Pb      {\ensuremath{\mathrm{b}}\xspace}                 
 \def\Pc      {\ensuremath{\mathrm{c}}\xspace}                 
 \def\Pd      {\ensuremath{\mathrm{d}}\xspace}                 
 \def\Pe      {\ensuremath{\mathrm{e}}\xspace}

 \def\Pi      {\ensuremath{\mathrm{i}}\xspace}

 \def\Pq      {\ensuremath{\mathrm{q}}\xspace}                 
                  
 \def\Ps      {\ensuremath{\mathrm{s}}\xspace}                 
                  
 \def\Pu      {\ensuremath{\mathrm{u}}\xspace}

 \def\thebaroffset{0.0em}
}
{

 \def\Pmu         {\ensuremath{\mu}\xspace}                 
 \def\Pnu         {\ensuremath{\nu}\xspace}

 \def\Ptau        {\ensuremath{\tau}\xspace}

 \def\Ppsi        {\ensuremath{\psi}\xspace}                 
                  
 \mathchardef\PDelta="7101
 \mathchardef\PXi="7104
 \mathchardef\PLambda="7103
 \mathchardef\PSigma="7106
 \mathchardef\POmega="710A
 \mathchardef\PUpsilon="7107
                  
 \def\PB      {\ensuremath{B}\xspace}                 
                  
 \def\PD      {\ensuremath{D}\xspace}

 \def\PJ      {\ensuremath{J}\xspace}                 
 \def\PK      {\ensuremath{K}\xspace}

 \def\PW      {\ensuremath{W}\xspace}

 \def\PZ      {\ensuremath{Z}\xspace}                 
                  
 \def\Pb      {\ensuremath{b}\xspace}                 
 \def\Pc      {\ensuremath{c}\xspace}                 
 \def\Pd      {\ensuremath{d}\xspace}                 
 \def\Pe      {\ensuremath{e}\xspace}

 \def\Pi      {\ensuremath{i}\xspace}

 \def\Pq      {\ensuremath{q}\xspace}                 
                  
 \def\Ps      {\ensuremath{s}\xspace}                 
                  
 \def\Pu      {\ensuremath{u}\xspace}

 \def\thebaroffset{0.18em}
}
\newcommand{\offsetoverline}[2][\thebaroffset]{\kern #1\overline{\kern -#1 #2}}%

\makeatletter
\ifcase \@ptsize \relax
  \newcommand{\miniscule}{\@setfontsize\miniscule{4}{5}}
\or
  \newcommand{\miniscule}{\@setfontsize\miniscule{5}{6}}
\or
  \newcommand{\miniscule}{\@setfontsize\miniscule{5}{6}}
\fi
\makeatother

\DeclareRobustCommand{\optbar}[1]{\shortstack{{\miniscule (\rule[.5ex]{1.25em}{.18mm})}
  \\ [-.7ex] $#1$}}

\def\electron   {{\ensuremath{\Pe}}\xspace}

\def\epem       {{\ensuremath{\Pe^+\Pe^-}}\xspace}

\def\muon       {{\ensuremath{\Pmu}}\xspace}
\def\mup        {{\ensuremath{\Pmu^+}}\xspace}

\def\mumu       {{\ensuremath{\Pmu^+\Pmu^-}}\xspace}

\def\tauon      {{\ensuremath{\Ptau}}\xspace}

\def\lepton     {{\ensuremath{\ell}}\xspace}

\def\neu        {{\ensuremath{\Pnu}}\xspace}

\def\W      {{\ensuremath{\PW}}\xspace}
\def\Wp     {{\ensuremath{\PW^+}}\xspace}
\def\Wm     {{\ensuremath{\PW^-}}\xspace}

\def\Z      {{\ensuremath{\PZ}}\xspace}

\def\quark     {{\ensuremath{\Pq}}\xspace}
\def\quarkbar  {{\ensuremath{\overline \quark}}\xspace}
\def\qqbar     {{\ensuremath{\quark\quarkbar}}\xspace}
\def\uquark    {{\ensuremath{\Pu}}\xspace}

\def\dquark    {{\ensuremath{\Pd}}\xspace}

\def\squark    {{\ensuremath{\Ps}}\xspace}

\def\cquark    {{\ensuremath{\Pc}}\xspace}
\def\cquarkbar {{\ensuremath{\overline \cquark}}\xspace}

\def\bquark    {{\ensuremath{\Pb}}\xspace}
\def\bquarkbar {{\ensuremath{\overline \bquark}}\xspace}
\def\bbbar     {{\ensuremath{\bquark\bquarkbar}}\xspace}

\def\KorKbar {\kern \thebaroffset\optbar{\kern -\thebaroffset \PK}{}\xspace}

\def\D       {{\ensuremath{\PD}}\xspace}

\def\DorDbar {\kern \thebaroffset\optbar{\kern -\thebaroffset \PD}\xspace}

\def\Dp      {{\ensuremath{\D^+}}\xspace}
\def\Dm      {{\ensuremath{\D^-}}\xspace}

\def\DpDm    {\ensuremath{\Dp {\kern -0.16em \Dm}}\xspace}

\def\B       {{\ensuremath{\PB}}\xspace}

\def\BorBbar {\kern \thebaroffset\optbar{\kern -\thebaroffset \PB}\xspace}

\def\Bd      {{\ensuremath{\B^0}}\xspace}

\def\BdorBdbar {\kern \thebaroffset\optbar{\kern -\thebaroffset \Bd}\xspace}

\def\Bs      {{\ensuremath{\B^0_\squark}}\xspace}

\def\BsorBsbar {\kern \thebaroffset\optbar{\kern -\thebaroffset \Bs}\xspace}

\def\jpsi     {{\ensuremath{{\PJ\mskip -3mu/\mskip -2mu\Ppsi}}}\xspace}

\def\Y#1S{\ensuremath{\PUpsilon{(#1S)}}\xspace}

\def\LorLbar     {\kern \thebaroffset\optbar{\kern -\thebaroffset \PLambda}\xspace}

\def\BF         {{\ensuremath{\mathcal{B}}}\xspace}
\def\BR         {\BF}

\def\to                 {\ensuremath{\rightarrow}\xspace}

\def\eps   {{\ensuremath{\varepsilon}}\xspace}

\def\AT#1     {\ensuremath{A_{\mathrm{T}}^{#1}}\xspace}           

\def\C#1      {\ensuremath{\mathcal{C}_{#1}}\xspace}                       
\def\Cp#1     {\ensuremath{\mathcal{C}_{#1}^{'}}\xspace}                    
\def\Ceff#1   {\ensuremath{\mathcal{C}_{#1}^{\mathrm{(eff)}}}\xspace}        
\def\Cpeff#1  {\ensuremath{\mathcal{C}_{#1}^{'\mathrm{(eff)}}}\xspace}       
\def\Ope#1    {\ensuremath{\mathcal{O}_{#1}}\xspace}                       
\def\Opep#1   {\ensuremath{\mathcal{O}_{#1}^{'}}\xspace}                    

\newcommand{\nospaceunit}[1]{\ensuremath{\text{#1}}}       
\newcommand{\aunit}[1]{\ensuremath{\text{\,#1}}}       

\newcommand{\tev}{\aunit{Te\kern -0.1em V}\xspace}
\newcommand{\gev}{\aunit{Ge\kern -0.1em V}\xspace}
\newcommand{\mev}{\aunit{Me\kern -0.1em V}\xspace}
\newcommand{\kev}{\aunit{ke\kern -0.1em V}\xspace}
\newcommand{\ev}{\aunit{e\kern -0.1em V}\xspace}
 
\newcommand{\mevc}{\ensuremath{\aunit{Me\kern -0.1em V\!/}c}\xspace}
\newcommand{\gevc}{\ensuremath{\aunit{Ge\kern -0.1em V\!/}c}\xspace}
\newcommand{\mevcc}{\ensuremath{\aunit{Me\kern -0.1em V\!/}c^2}\xspace}
\newcommand{\gevcc}{\ensuremath{\aunit{Ge\kern -0.1em V\!/}c^2}\xspace}

\def\mum  {\ensuremath{\,\upmu\nospaceunit{m}}\xspace}

\def\barn{\aunit{b}\xspace}

\def\fb   {\ensuremath{\aunit{fb}}\xspace}
\def\invfb   {\ensuremath{\fb^{-1}}\xspace}

\def\ps   {\ensuremath{\aunit{ps}}\xspace}

\def\gsim{{~\raise.15em\hbox{$>$}\kern-.85em
          \lower.35em\hbox{$\sim$}~}\xspace}
\def\lsim{{~\raise.15em\hbox{$<$}\kern-.85em
          \lower.35em\hbox{$\sim$}~}\xspace}

\def\pt         {\ensuremath{p_{\mathrm{T}}}\xspace}

\def\ptot       {\ensuremath{p}\xspace}

\def\evtgen     {\mbox{\textsc{EvtGen}}\xspace}

\def\geant      {\mbox{\textsc{Geant4}}\xspace}

\def\photos     {\mbox{\textsc{Photos}}\xspace}

\def\pythia     {\mbox{\textsc{Pythia}}\xspace}

\def\tell1  {TELL1\xspace}
\def\ukl1   {UKL1\xspace}

\usepackage{cite} 
\usepackage{mciteplus}

\begin{document}
\renewcommand{\thefootnote}{\fnsymbol{footnote}}
\setcounter{footnote}{1}


\begin{titlepage}
\pagenumbering{roman}

\vspace*{-1.5cm}
\centerline{\large EUROPEAN ORGANIZATION FOR NUCLEAR RESEARCH (CERN)}
\vspace*{1.5cm}
\noindent
\begin{tabular*}{\linewidth}{lc@{\extracolsep{\fill}}r@{\extracolsep{0pt}}}
\ifthenelse{\boolean{pdflatex}}
{\vspace*{-1.5cm}\mbox{\!\!\!\includegraphics[width=.14\textwidth]{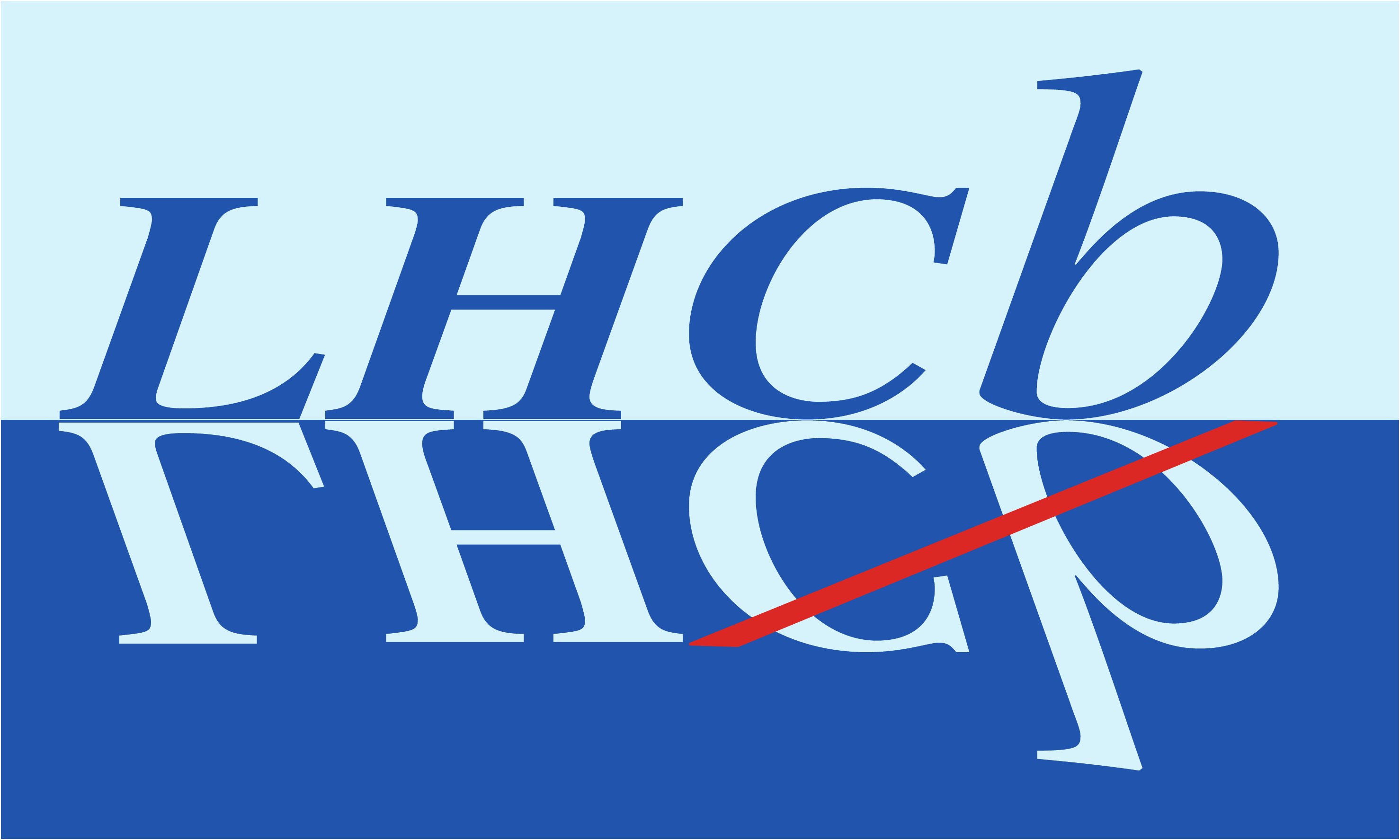}} & &}%
{\vspace*{-1.2cm}\mbox{\!\!\!\includegraphics[width=.12\textwidth]{figs/lhcb-logo.eps}} & &}%
\\
 & & CERN-EP-2020-194 \\  
 & & LHCb-PAPER-2020-022 \\  
 & & March 25, 2021 \\ 
 & & \\
\end{tabular*}

\vspace*{4.0cm}

{\normalfont\bfseries\boldmath\huge
\begin{center}
  \papertitle 
\end{center}
}

\vspace*{2.0cm}

\begin{center}
LHCb collaboration\paperauthors\footnote{Authors are listed at the end of this paper.}
\end{center}

\vspace{\fill}

\begin{abstract}
  \noindent
  A search is performed for heavy neutrinos in the decay of a $W$ boson into two muons and a jet. 
  The data set corresponds to an integrated luminosity of approximately $3.0\invfb$ of proton-proton collision data at centre-of-mass energies of 7 and $8\tev$ collected with the LHCb experiment.
  Both same-sign and opposite-sign muons in the final state are considered.
Data are found to be consistent with the expected background.
  Upper limits on the coupling of a heavy neutrino with the Standard Model neutrino are set at $95\%$ confidence level in the heavy-neutrino mass range from 5 to $50\gevcc$. These are of the order of $10^{-3}$ for lepton-number-conserving decays and of the order of $10^{-4}$ for lepton-number-violating heavy-neutrino decays.
\end{abstract}

\vspace*{2.0cm}

\begin{center}
  Published in
  Eur.~Phys.~J.~C~81~(2021)~248 
\end{center}

\vspace{\fill}

{\footnotesize 
\centerline{\copyright~\papercopyright. \href{\paperlicenceurl}{\paperlicence}.}}
\vspace*{2mm}

\end{titlepage}


\newpage
\setcounter{page}{2}
\mbox{~}


\renewcommand{\thefootnote}{\arabic{footnote}}
\setcounter{footnote}{0}

\cleardoublepage


\pagestyle{plain} 
\setcounter{page}{1}
\pagenumbering{arabic}


\section{Introduction}
\label{sec:introduction}

\newcommand{\tevcc}{\ensuremath{\aunit{Te\kern -0.1em V\!/}c^2}\xspace}

Many theories beyond the Standard Model (SM) predict the existence of heavy neutral leptons (HNLs) to explain the smallness of neutrino masses\cite{PhysRevLett.44.912,Yanagida:1979as,GellMann:1980vs}. 
These leptons, $N$, could be observed at collider experiments if their masses are at the electroweak scale.
The HNLs may mix with the light SM neutrinos $\nu_\ell$, with a strength
governed by the coupling $V_{N\ell}$. 
The mixing matrix is not expected to be flavour diagonal, which leads to signatures with transitions between different lepton flavours.
Experimentally, direct searches for a generic heavy neutrino are performed through their mixing with each flavour of SM neutrino, typically considering decays where no flavour mixing occurs.
The HNL is expected to be long-lived if the coupling is small enough.
This analysis searches for the mixing of a heavy neutrino with a muon neutrino, taking advantage of the high reconstruction efficiency for muons at LHCb. 
The HNL mass range covered is between \SIlist{5;50}{\gevcc}.
The dominant HNL production mechanism in this mass range is via the decay of gauge bosons, $\W^{\pm}\to\lepton^{\pm}\neu$ and $\Z\to\neu \neu$, where one of the SM neutrinos mixes with the heavy neutrino.
For brevity, the processes $\W^{\pm}\to\lepton^{\pm}\neu [\neu\to N]$ and $\Z\to\neu \neu [\neu\to N]$ will be written as $\W^{\pm}\to\lepton^{\pm}N$ and $\Z\to\neu N$ throughout.
Both lepton-number-violating and lepton-number-conserving decays of a heavy neutrino are considered.
The heavy neutrino is assumed to have negligibly small lifetime. 

The DELPHI collaboration was first to set a limit on these types of decays considering $\Z\to\nu N$ decays in \epem{} collisions at the \Z{} resonance, where both long- and short-lived signatures were analysed~\cite{Abreu:1996pa}. 
The upper limit on the $\Z\to\nu N$ branching fraction of \num{1.3e-6} at 95\% confidence level (CL) for $N$ masses between \SIlist{3.5;50}{\gevcc} leads to one of the most stringent constraints on the coupling in this mass range.
At the LHC, a more promising detection approach for $N$ with mass below the weak scale are leptonic decays of \W{} bosons, $\W^\pm\to \lepton^{\pm} N$.
Searches by the ATLAS~\cite{Aad:2011vj,ATLAS:2012ak,Aad:2015xaa,Aaboud:2018spl,Aaboud:2019wfg} and CMS~\cite{Sirunyan:2018vhk,Sirunyan:2018pom,Khachatryan:2016olu,Khachatryan:2015gha,Khachatryan:2014dka,CMS:2012zv,Chatrchyan:2012fla} collaborations at centre-of-mass energies of 7, 8 and $13\tev$ typically probed larger neutrino masses, from $40\gevcc$ up to $2700\gevcc$, employing a signature of two same-sign leptons and two jets.   
A recent search performed by the CMS collaboration also included final states with at least one jet, extending the probed heavy-neutrino mass range down to $20\gevcc$~\cite{Sirunyan:2018xiv}.
In the mass range studied in this analysis, searches of promptly decaying heavy neutrinos in leptonic final states of the \W boson at centre-of-mass energy of $13 \tev$ by the ATLAS~\cite{Aad:2019kiz} and CMS~\cite{Sirunyan:2018mtv} collaborations set constraints comparable to that of the DELPHI collaboration. 
A long-lived signature has also been explored by the ATLAS collaboration, excluding coupling strengths down to about $10^{-6}$ between 4.5 and $10\gevcc$, and hence representing the most stringent limit to date in this mass range~\cite{Aad:2019kiz}.

The branching fraction (\BF) for the decay of a \W boson into a muon and a heavy neutrino is suppressed with respect to the SM decay $\W^+\to\muon^+\neu$ by the mixing of the active neutrino with the heavy neutrino and a phase-space factor according to Ref.~\cite{DITTMAR19901}
\begin{equation}
  \BF(\W\to \muon N) = \BF(\W\to \muon \neu)\left| V_{\muon N}\right|^2\left(1-\frac{m_N^2}{m_W^2}\right)^{2}\left(1+\frac{m_N^2}{2m_W^2}\right).
  \label{equ:WBR}
\end{equation}
The heavy neutrino decays via neutral or charged current interactions $N \to \nu Z^{(*)}$, $N \to \nu H^{(*)}$ or $ N \to \muon^{\pm}\W^{\mp(*)}$, where the \Z, Higgs and \W bosons can be on- or off-shell. 
The corresponding branching fractions are computed based on Refs.~\cite{Helo:2011yg,Helo:2010cw}, where the Higgs contribution is neglected due to its suppression in the mass range considered. 
The total width is given by the sum of the partial decay widths of charged and neutral current interactions. %
If the neutrino is a Majorana particle, an additional lepton-number-violating decay contributes to the same final state, with the same partial decay width as the lepton-number-conserving decay. 
The branching fraction to any non-charge-specific final state is unaffected, but the lifetime is a factor of two smaller than if the neutrino were a Dirac particle.
 
\begin{figure}[t]
  \begin{center}
    {\includegraphics[width=0.48\textwidth]{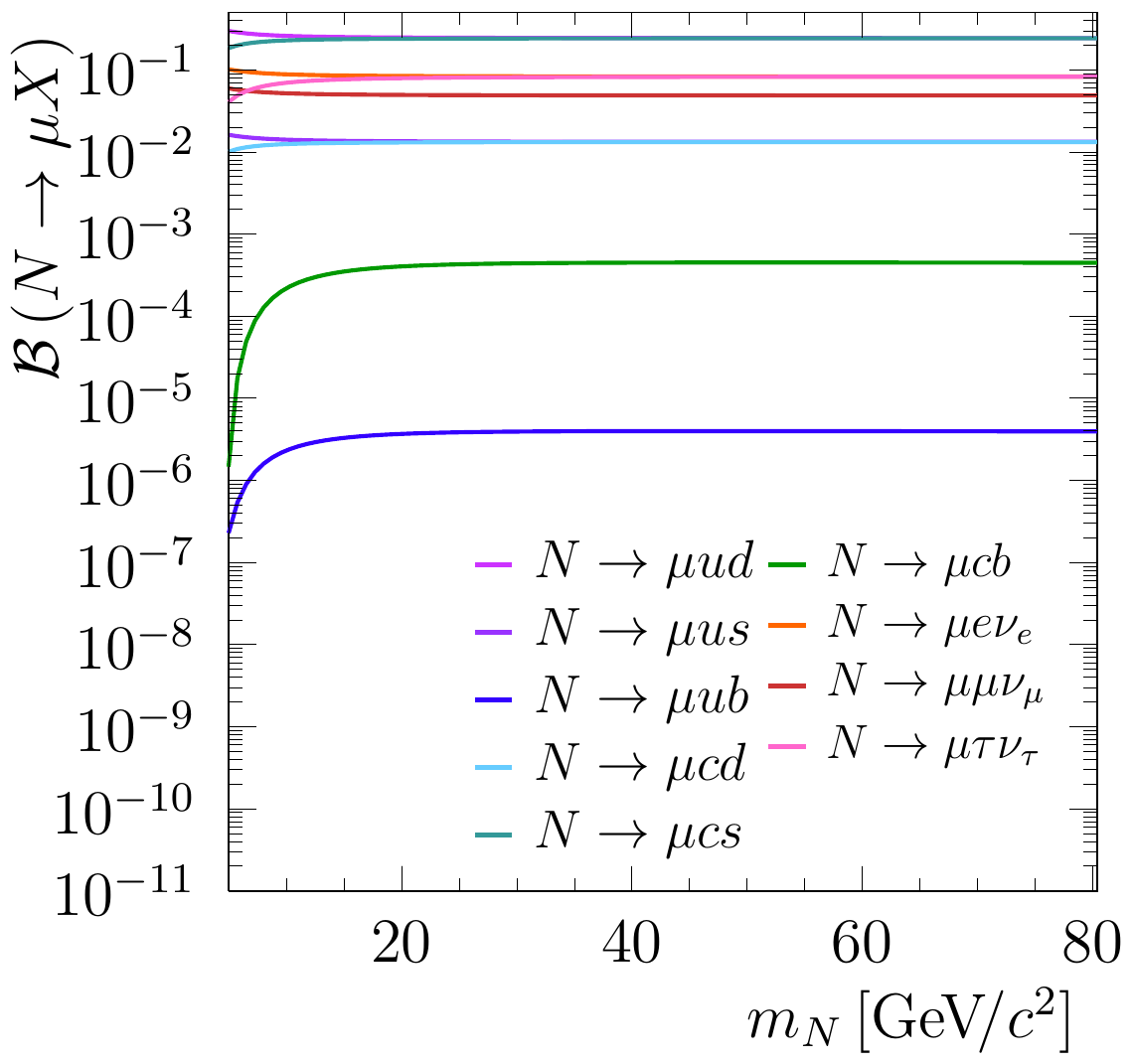}}\hspace{2mm}
    {\includegraphics[width=0.48\textwidth]{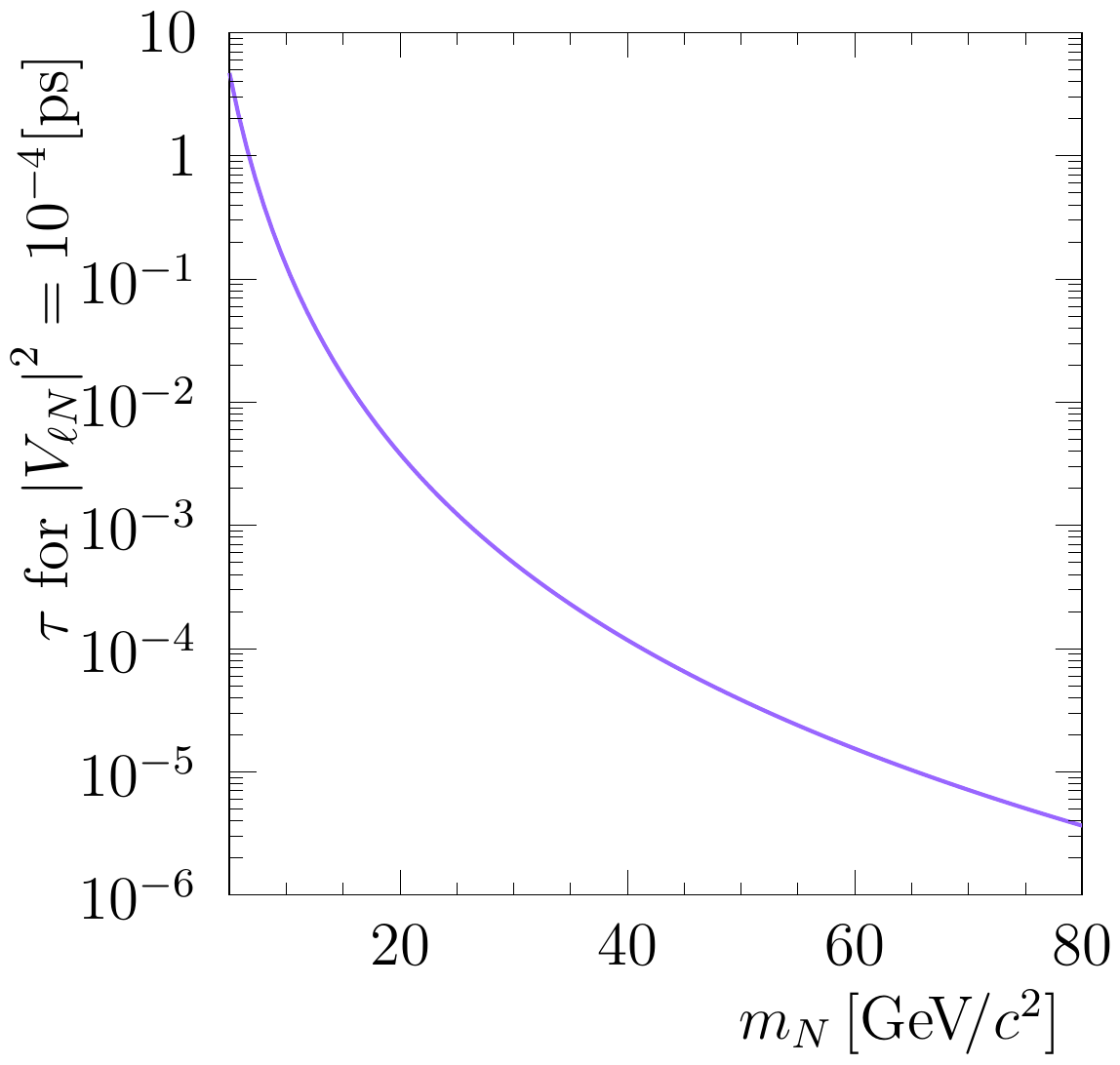}}
  \end{center}
  \caption{Properties of a heavy neutrino as a function of its mass~\cite{Helo:2011yg,Helo:2010cw}: (left) the branching fractions to final states with a muon and (right) the lifetime, assuming a coupling of $10^{-4}$.}
  \label{fig:HNBR}
\end{figure}

The left plot of \fig\ref{fig:HNBR} shows the branching fraction for HNL decay modes with a muon in the final state as a function of the heavy-neutrino mass.
The difference between the HNL decay modes to quarks is mainly due to CKM matrix elements~\cite{Cabibbo:1963yz,Kobayashi:1973fv}, with the quark masses only playing a minor role at low heavy-neutrino masses.
The branching fraction of the decay $N\to \muon\muon\neu$ is about one order of magnitude smaller than that of the $N\to\muon\qqbar'$ mode, due to negative interference between charged and neutral current interactions.
In the right plot of \fig\ref{fig:HNBR} the lifetime is shown as a function of the heavy-neutrino mass assuming a coupling of $10^{-4}$. 
In the low-mass regime, the lifetime is of the order of a few $\ps$, while at higher masses the lifetime is so small that the decay can be considered prompt.

\begin{figure}[t!]
  \centerline{\includegraphics[width=0.68\textwidth]{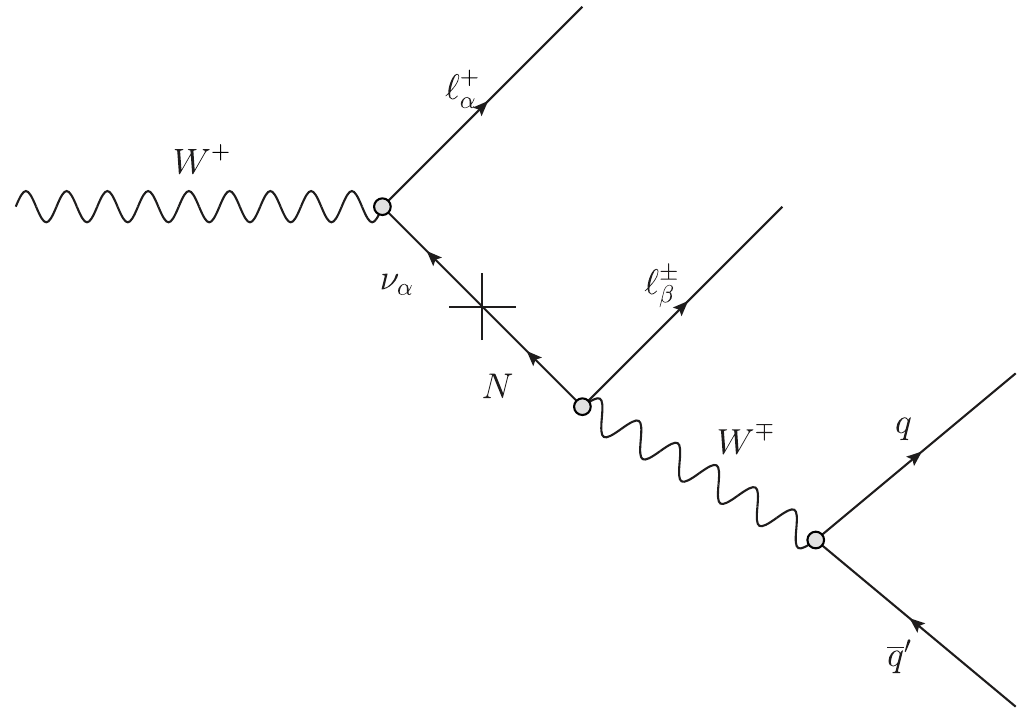}}
  \caption{Feynman diagram for the production of a heavy neutrino via
    mixing with a neutrino from the decay of a \W boson and
    semileptonic decay of the heavy neutrino into a lepton and two
    quarks. The subscripts $\alpha$ and $\beta$ indicate the lepton flavour. In this analysis $\alpha$ and $\beta$ are both muons.}
  \label{fig:feynman}
\end{figure}

In this paper, a search is presented for a prompt HNL in the decay\footnote{Charge-conjugate processes are implied throughout the paper.} $\Wp\to\mup N$ with $N\to\mu^{\pm}\quark\quarkbar'$, as depicted in \fig\ref{fig:feynman}.
Data collected by the LHCb experiment in proton-proton collisions at centre-of-mass energies of \SI{7}{\tev} in 2011 and \SI{8}{\tev} in 2012 are used, corresponding to integrated luminosities of \SIlist[per-mode=reciprocal]{1.0;1.9}{\invfb}~\cite{LHCb-PAPER-2014-047}, respectively.

The experimental signature consists of two muons and one or two jets depending on the HNL mass.
The muon from \W decay, denoted as \muonW, carries significant transverse momentum, while the muon from $N$ decay, denoted as \muonN, has lower momentum.
Both same-sign and opposite-sign muons are considered, allowing for the possibility that the HNL has a Majorana nature.
The signal yields for both categories and several mass hypotheses in the range $5-50\gevcc$ are extracted from the data and normalized with respect to the $\Wp\to\mup\neu$ decay. Corresponding upper limits are then set on the product of coupling and branching ratio.

The paper is organised as follows.
In \sect\ref{sec:detector} the detector, data and simulation samples are described, and in \sect\ref{sec:selection} the selection of signal and normalisation candidates is discussed. 
\sect\ref{sec:results} contains the results and conclusions are drawn in \sect\ref{sec:conclusion}.

\section{Detector and simulation}
\label{sec:detector}

The \lhcb detector~\cite{LHCb-DP-2008-001,LHCb-DP-2014-002} is a
single-arm forward spectrometer covering the \mbox{pseudorapidity}
range $2<\eta <5$, designed for the study of particles containing
\bquark or \cquark quarks. The detector includes a high-precision
tracking system consisting of a silicon-strip vertex detector
surrounding the $pp$ interaction
region~\cite{LHCb-DP-2014-001}, a large-area silicon-strip
detector located upstream of a dipole magnet with a bending power of
about $4{\mathrm{\,Tm}}$, and three stations of silicon-strip
detectors and straw drift
tubes~\cite{LHCb-DP-2013-003}
  placed
downstream of the magnet.  The tracking system provides a measurement
of the momentum, \ptot, of charged particles with a relative
uncertainty that varies from 0.5\% at low momentum to 1.0\% at
200\gevc.  The minimum distance of a track to a primary proton-proton collision vertex (PV),
the impact parameter (IP), is measured with a resolution of
$(15+29/\pt)\mum$, where \pt is the component of the momentum
transverse to the beam, in\,\gevc.  Different types of charged hadrons
are distinguished using information from two ring-imaging Cherenkov
detectors~\cite{LHCb-DP-2012-003}.  Photons, electrons and
hadrons are identified by a calorimeter system consisting of
scintillating-pad (SPD) and preshower detectors, an electromagnetic
and a hadronic calorimeter. Muons are identified by a
system composed of alternating layers of iron and multiwire
proportional chambers~\cite{LHCb-DP-2012-002}.
The online event selection is performed by a
trigger~\cite{LHCb-DP-2012-004}, which consists of a hardware stage,
based on information from the calorimeter and muon systems, followed
by a software stage, which applies a full event reconstruction. For
the events selected for this analysis, the trigger requires at least a single
muon with $\pt{}>1.48\, (1.76)\gevc$ at the hardware stage in 2011 (2012), and includes an upper
threshold of 600 hits in the SPD to prevent high-particle multiplicity
events from dominating the processing time. A muon with $\pt{}>10\gevc$
is required at the software stage.

Simulated samples were generated for the signal decay with both opposite- and same-sign muons in the final state, in equal amount.
Samples were generated for HNL masses of 5, 10, 15, 20, 30, and $50\gevcc$, using the minimal mixing scenario model~\cite{Batell:2016zod} and accounting for angular correlations due to spin effects. 
The parton level process is generated
with \madgraph
5~\cite{Alwall:1699128,Frederix:2018nkq}, while \pythia 8~\cite{Sjostrand:2007gs},
with a specific \lhcb configuration~\cite{LHCb-PROC-2010-056}, is used
for the generation of the underlying event, fragmentation and
hadronisation. Decays of hadronic particles are described by
\evtgen~\cite{Lange:2001uf}, in which final-state radiation is
generated using \photos~\cite{Golonka:2005pn}. The interaction of the
generated particles with the detector, and its response, are
implemented using the \geant toolkit~\cite{Allison:2006ve,
  *Agostinelli:2002hh} as described in Ref.~\cite{LHCb-PROC-2011-006}.
Simulated background samples are generated using \pythia 8.
The \textsc{DYTurbo}~\cite{Camarda:2019zyx} program is used to correct the kinematic distributions of the simulated $\Wp\to\mup\neu$ background.

\section{Event selection}
\label{sec:selection}

Signal candidates are reconstructed from a pair of charged tracks identified as muons and a single jet.
First, the high-momentum muon, \muonW, is selected. 
Both the hardware and software trigger decisions are required to be associated to the high-momentum muon candidate. 
The track is required to have transverse momentum larger than $20\gevc$, to be of good quality and to have a high significance of the track curvature to remove high transverse momentum tracks with poorly determined charge.
The high-momentum muon candidate is also required to have small relative energy deposition in the calorimeters to reject pions and kaons 
misidentified as muons. 
The muon selection criteria for the normalisation channel $\Wp\to\mup\neu$ are the same as for the high-momentum muon of the signal.

The lower-momentum muon candidate, \muonN, is required to have transverse momentum higher than $3\gevc$.
The combined invariant mass of the \muonN and \muonW candidates must be in the range \num{20} to \SI{70}{\gevcc} to suppress the background from $\Z\to\mu\mu$ decays.
Depending on the relative charge of the two muons the candidates are classified as same-sign (SS) or opposite-sign (OS).

Jets are reconstructed following a particle flow approach~\cite{LHCb-PAPER-2013-058}, using tracks of charged particles and calorimeter energy deposits as inputs. 
To prevent overlap between jets and signal muons, tracks identified as muons with a transverse momentum greater than $2\gevc$ are excluded from the jet reconstruction. 
The anti-$k_T$ jet clustering algorithm is used~\cite{Cacciari:2008gp}, with a distance parameter $R = \sqrt{(\Delta\phi)^2+(\Delta\eta)^2}= 0.5$, where $\phi$ is the azimuthal angle and $\theta$ the pseudorapidity.
The jet four-momentum is calculated from the four-vectors of its constituents, and corrected for pollution from pile-up and the underlying event using the per-event particle multiplicity~\cite{LHCb-PAPER-2013-058}.
To enhance the jet purity the fraction of the jet energy carried by charged particles should be at least 0.1, the jet must have $\pt>10\gevc$ and contain at least one track with $\pt>1.2\gevc$.
Only candidates with at least one jet passing these criteria are retained.
Jets are combined with lower-momentum muon candidates to form $N\to\muonN\jet$ candidates, which are required to have invariant mass smaller than \SI{80}{\gevcc} and a transverse momentum greater than $10\gevc$. 
The selected heavy-neutrino candidates are then combined with a high-momentum muon candidate to form \W candidates.
Since the assignment of the two muons is ambiguous if they both satisfy the high-momentum muon selection, the mass, $m(\muonN\jet)$, of the $\muonN \jet$ combination is required to be smaller than that of the $\muonW \jet$ combination.
Only the $\muonW \muonN \jet$ candidates within \SI{20}{\gevcc} of the known \W mass~\cite{Tanabashi:2018oca} are retained.

A scale factor is applied to the jet four-momentum, constraining the invariant mass of the $\muonW\muonN\jet$ system to the known mass of the \W boson.
This leads to a significant improvement in the resolution of $m(\muonN\jet)$ and  diminishes the sensitivity of the heavy-neutrino mass distribution to the jet energy scale.

Dominant background sources are charged weak currents, in particular $pp\to W + X $ with $\W \to \mu \nu$ or $\W \to\tau \nu$, neutral electroweak Drell-Yan processes $pp \to \gamma/Z^{(*)}+X$ with $\gamma/\Z^{(*)} \to \mu \mu, \tau\tau$, heavy flavour $\bquark\bquarkbar \to X\mu$ and $\cquark\cquarkbar \to X \mu$, and $X\muon$ production from light QCD $(\uquark,\dquark,\squark)$.
In the same-sign muon channel the Drell-Yan type background contributions are highly suppressed, while in the opposite-sign muon channel the contribution from low-mass Drell-Yan processes remains a prominent irreducible background.

Most of the heavy flavour background is suppressed by requiring the IP for \muonW and \muonN to be less than $40\mum$ and $100\mum$, respectively.
The remaining background is reduced by using three different multivariate classifiers based on a Boosted Decision Tree (BDT) algorithm~\cite{Breiman:2253780,Roe:2004na,FREUND1997119}. 
The three classifiers are referred to as the \muonW{} uBDT, the \muonN{} uBDT and the kinematics uBDT: the first two classifiers are dedicated to the identification of the respective muons, while the latter exploits the event kinematics to distinguish the signal from the remaining background.
All three are trained minimising the dependence of the signal efficiency on the true neutrino mass using the uBoost method~\cite{Stevens:2013dya}.
The training of all classifiers uses a cross-validation technique~\cite{Blum:1999:BHB:307400.307439}.
The classifiers are trained using simulated decays of the heavy neutrino with same-sign muons in the final state as a proxy for signal.
Both charged and neutral weak background contributions have a muon in the final state with similar kinematics to the signal high-momentum muon. 
The \muonW classifier discriminates between the signal and heavy flavour background.
It is trained using data candidates where both muons have large impact parameters ($\text{IP}(\muonW)>\text{\SI{40}{\mum}}$, $\text{IP}(\muonN)>\text{\SI{100}{\mum}}$) as a proxy for background.
For both the \muonN uBDT and the kinematics uBDT,  a combination of the dominant background sources from simulation is used.
The input variables used for each of the muon identification classifiers are the combined particle identification information from the RICH, calorimeter and muon systems, the ratio of the energy deposited in both calorimeters to the measured track momentum, and observables describing the isolation of the tracks.
Additional isolation variables of different cone sizes are included among the inputs for the \muonN uBDT classifier.
The input variables of the kinematics classifier comprise the angular distance $R$ between the \muonN and the jet, the angle between the two muons in the rest frame of the heavy neutrino, the transverse component of the sum of the four-momentum of all particles used as particle flow input, the dimuon mass, the combined invariant mass of the dimuon and the jet, and the jet transverse momentum.
The optimal requirement on the output of each BDT classifier is selected by maximising the Punzi figure-of-merit~\cite{Punzi:2003bu} for three units of significance.
This is first evaluated for the \muonW uBDT, followed by the simultaneous optimisation of the \muonN and kinematics uBDTs. 
The optimal requirements are found to be the same for all the simulated signal samples.
The selection is optimised for the same-sign muon signal, but it is verified to be optimal for the opposite-sign category as well, since the differences in spin-dependent observables between the two channels have a negligible effect on the output distributions of the BDT classifiers. 

The background sources are studied and evaluated in three control regions: one enhanced in electroweak \W background components, one in heavy flavour background components and one in light QCD background components, indicated as \W, \bquark\bquarkbar and QCD regions, respectively.
The requirements defining the control regions with respect to the signal region are reported in \tab\ref{tab:bkgcuts}.
An additional region, denoted as the $\Z\to\muon\muon$ region, is defined by the following criteria: both muons are required to have transverse momentum greater than $20\gevc$ and IP smaller than \SI{40}{\mum} and the invariant mass of the muon pair must be between \SIlist{60;120}{\gevcc}. 
In each control region the predicted background composition and yield are compared to the data to confirm that no other contribution has been neglected.

\begin{table}[tb]
  \caption{Requirements on IP and BDT classifiers defining the signal and control regions.}
  \label{tab:bkgcuts}
  \centerline{
    \begin{tabular}{lccccc}
      \toprule
                                 & $\text{IP}(\muonW)$ [mm] & \muonW uBDT & \muonN uBDT & Kinematic uBDT&       $\text{IP}(\muon_{N})$ [mm] \\
      \midrule
      Signal                     & $< 0.04$ & $ > 0.55$ & $ > 0.60 $ & $ > 0.62 $ & $ < 0.1$ \\
      $\W$ region                & $< 0.04$ & $ > 0.55$ & $ < 0.60 $ & $ < 0.62 $ & $ < 0.1$ \\
      $\bquark\bquarkbar$ region & $> 0.04$ & $ < 0.55$ & $ < 0.60 $ & $ < 0.62 $ & $ > 0.1$ \\
      QCD region                 & $< 0.04$ & $ < 0.55$ & $ > 0.60 $ & $ > 0.62 $ & $ < 0.1$ \\
      \bottomrule
    \end{tabular}
  }
\end{table}

\section{Fit strategy and results}
\label{sec:results}

\newcommand{\Nsig}{\ensuremath{N_{\text{sig}}}}
\newcommand{\Nnorm}{\ensuremath{N_{\text{norm}}}}

The product of the branching fraction $\BR(N\to \muon \jet)$ and the squared coupling $\left|V_{\muon N}\right|^2$ is proportional to the number of signal candidates, $N_{\text{sig}}$, and can be written with respect to the number of $\W\to\muon\neu$ candidates as 
\begin{equation}
  \BR(N\to \muon \jet) \left| V_{\muon N}\right|^2 = \frac{N_{\text{sig}}}{N_{\text{norm}}}\frac{\eps_{\text{norm}}}{\eps_{\text{sig}}}\left(1-\frac{m_N^2}{m_W^2}\right)^{-2}\left(1+\frac{m_N^2}{2m_W^2}\right)^{-1},
\end{equation}
where $N_{\text{sig}}$ and $N_{\text{norm}}$ denote the yields of the signal and normalisation channels and $\eps_{\text{sig}}$ and $\eps_{\text{norm}}$ their efficiencies.
The phase-space suppression factor and the coupling term arise from the heavy-neutrino production process described by \eq\ref{equ:WBR}. 
The $\Wp\to\mup\neu$ branching fraction in \eq\ref{equ:WBR} cancels with the normalisation channel.

\subsection{Normalisation channel}
\label{subsec:norm}
\begin{figure}[tb]
  \centering
  {\includegraphics[width=0.48\textwidth]{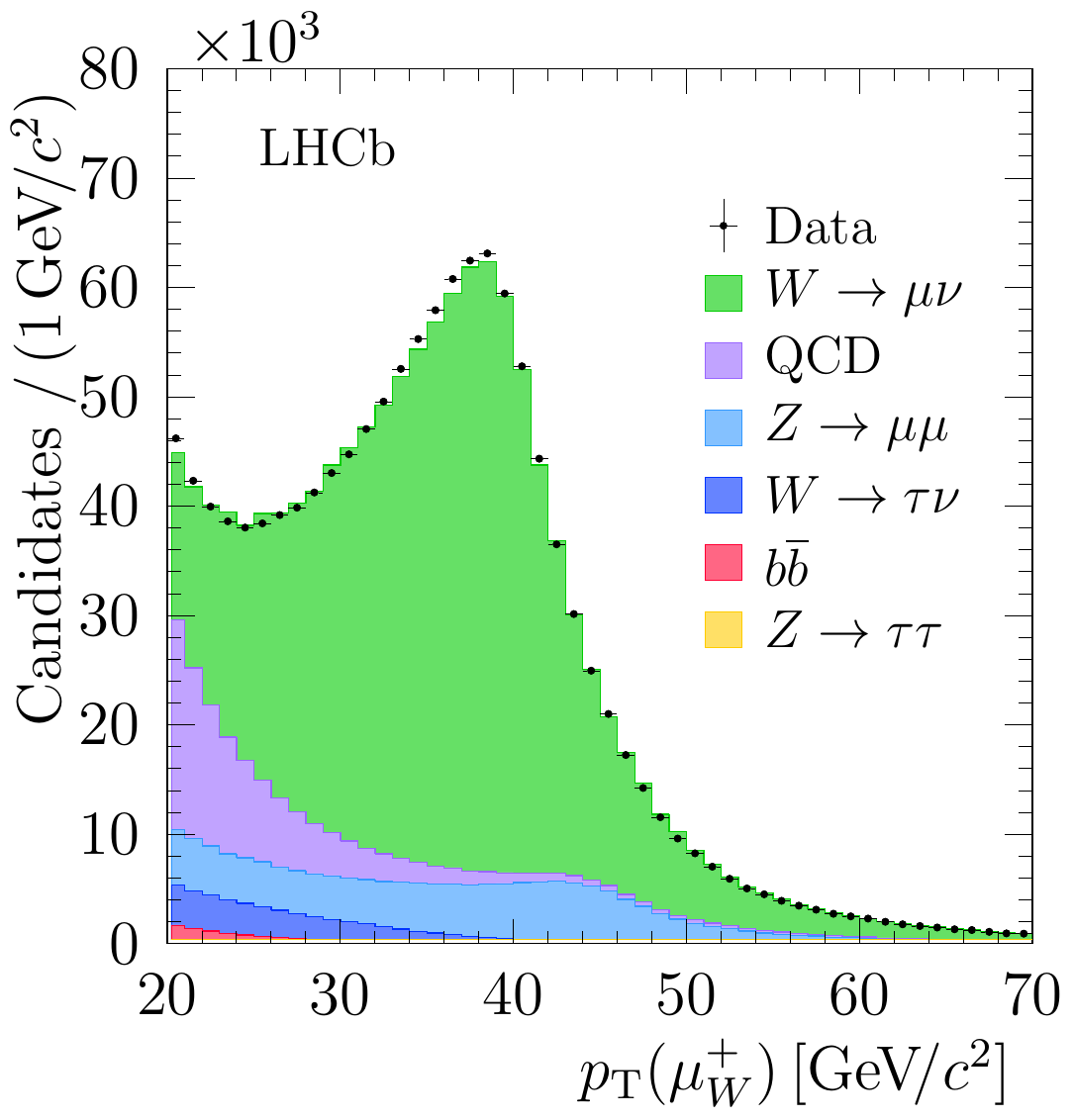}}\hspace{2mm}
  {\includegraphics[width=0.48\textwidth]{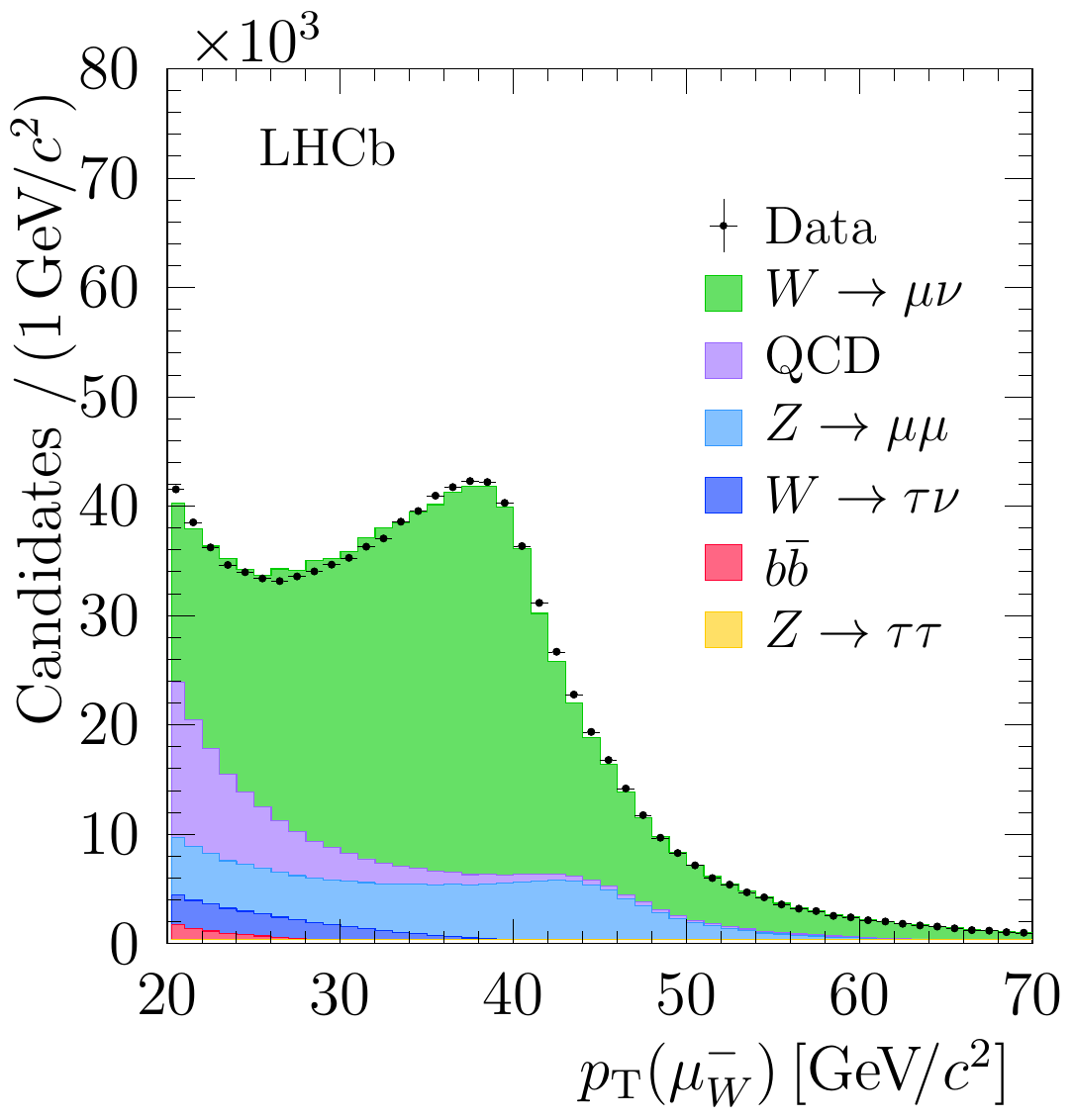}}
  \caption{The (left) positive and (right) negative muon transverse momentum spectra for the 2012 data set integrated over pseudorapidity for the normalisation channel. The filled histograms are the result of the fit to the data.}
    \label{fig:normfit}
  \end{figure}

The yield of the normalisation channel is determined using a binned maximum-likelihood fit to the muon transverse momentum distribution separately for each year of data taking and in eight bins of muon pseudorapidity.
The fit is performed separately for positively and negatively charged muons to account for the difference in production rate at LHCb. 
The main background contributions are $\gamma/\Z^{(*)} \to \mu   \mu$ decays and hadron misidentification (denoted as QCD).
Minor contamination from $\Z\to\tau\tau$, $\W\to\tau\nu$ and $\bbbar$ processes is also present. 
The templates are obtained in bins of pseudorapidity from simulation for each component, with the exception of the QCD background templates that are determined from a control sample characterised by large energy deposits in the calorimeters. 
The yields for the minor background contributions are fixed to their expected values from simulation. 
The yield for the $\Z\to\muon\muon$ component is constrained to the value obtained from the corresponding control region extrapolated according to simulation. 
The distribution of the muon transverse momentum for 2012 data integrated over pseudorapidity is shown in \fig\ref{fig:normfit} with the filled histograms resulting from the fit to the data overlaid.

Systematic uncertainties on the normalisation yield are estimated separately for the 2011 and 2012 data sets by varying the shape and normalisation of the templates.
Replacing the QCD template with an exponential distribution and 
varying the $\W\to\muon\neu$ templates each yield a difference with respect to the default fit of about 1\%, which is assigned as a systematic uncertainty.
The ratio of measured QCD yields per pseudorapidity bin between positively and negatively charged muons deviates from unity. 
A systematic uncertainty of 0.7\% is assigned to account for the difference with respect to the default fit when the normalisation of the QCD component is fixed bin by bin to the average of the yields.
Systematic uncertainties of less than 0.1\% are assigned for each of the components whose yield is fixed in the fit to account for the largest variation observed with respect to the default fit when each yield is changed by one standard deviation. 
For the 2011 data set an additional source of uncertainty is considered to account for the difference in templates between 2011 and 2012 simulation, resulting in 0.7\% assigned systematic uncertainty.
The total systematic uncertainty on the normalisation yield is 1.8\% and 1.6\% for the 2011 and 2012 data sets, respectively.

The total yield for the normalisation channel $\W\to\muon\neu$ is $(795\pm 1 \pm 15)\times 10^3$ for 2011 data and $(1719 \pm 2 \pm 27)\times 10^3$ for 2012 data, where the first uncertainty is statistical and the second systematic.
The total yield comprises 57\% \Wp decays and 43\% \Wm decays. 
The ratio of the measured yields for positively and negatively charged muons as a function of pseudorapidity is in good agreement with the simulation and the measurement of Ref.~\cite{LHCb-PAPER-2015-049}.

\subsection{Efficiency ratio}
\label{subsec:effratio}
The efficiency and corresponding uncertainties of the selection requirements for both the normalisation and signal samples are determined separately for each year of data taking using simulation.
Corrections to account for mismodelling in simulation are derived from control samples, such as $\Z\to\mumu$ and $\jpsi\to\mumu$, and are applied to the efficiencies related to the reconstruction of the two muons, the required number of hits in the SPD and the \muonW uBDT and \muonN uBDT criteria. 
When sufficient data is available, the corrections are evaluated in bins of pseudorapidity and momentum or transverse momentum of the muon.
The dominant source of systematic uncertainties arises from the different detector response to jets between simulation and data.
The energy scale is modelled to an accuracy of about 5\%, driven mainly by the response to neutral particles, while the jet energy resolution is modelled in simulation to an accuracy of about 10\%~\cite{LHCb-PAPER-2013-058,LHCb-PAPER-2014-055,LHCb-PAPER-2015-021}.
The corresponding systematic uncertainties on the efficiency ratio are evaluated in simulation by varying the jet energy by 5\% for the former and by smearing the jet transverse momentum by 10\% for the latter. 
Both resulting uncertainties vary between 5\% and 11\% depending on the heavy-neutrino mass, where the fluctuation is due to the limited size of the simulated samples. 
The overall uncertainty due to jet identification requirements, which  amounts to 1.7\%, is taken from Ref.~\cite{LHCb-PAPER-2016-011}.
A systematic uncertainty to account for the mismatch between simulation and data of the missing transverse momentum in the event varies between 1\% and 2.5\% depending on the heavy-neutrino mass.
The uncertainties related to the efficiency of the \muonW selection largely cancel for the signal and normalisation modes, since their selections are identical.
The relative uncertainty on the correction factors is of the order of 2\%.
The ratios of efficiencies between the normalisation and signal channel, for different heavy neutrino masses, are reported in \tab\ref{tab:ratio_eff_corr}.

\begin{table}[bt]
  \caption{Efficiency ratios, for different heavy-neutrino masses, between normalisation and signal channels. The first  uncertainty is statistical, the second is systematic.}
    \centerline{\begin{tabular}{c 
>{\collectcell\num}r<{\endcollectcell} @{${}\pm{}$} >{\collectcell\num}l<{\endcollectcell} @{${}\pm{}$} >{\collectcell\num}l<{\endcollectcell}
>{\collectcell\num}r<{\endcollectcell} @{${}\pm{}$} >{\collectcell\num}l<{\endcollectcell} @{${}\pm{}$} >{\collectcell\num}l<{\endcollectcell}
>{\collectcell\num}r<{\endcollectcell} @{${}\pm{}$} >{\collectcell\num}l<{\endcollectcell} @{${}\pm{}$} >{\collectcell\num}l<{\endcollectcell}
>{\collectcell\num}r<{\endcollectcell} @{${}\pm{}$} >{\collectcell\num}l<{\endcollectcell} @{${}\pm{}$} >{\collectcell\num}l<{\endcollectcell}
}
\toprule
  $N$ mass & \multicolumn{6}{c}{Same sign} & \multicolumn{6}{c}{Opposite sign}\\
  $[\gevcc]$ & \multicolumn{3}{c}{2011} & \multicolumn{3}{c}{2012} & \multicolumn{3}{c}{2011} & \multicolumn{3}{c}{2012}\\
\midrule
\phantom{0}5  & 24 & 1 & 2 & 25 & 1 & 3 & 22 & 1 & 2 & 21 & 1 & 1\\
10 & 24 & 1 & 2 & 24 & 1 & 2 & 21 & 1 & 2 & 19 & 1 & 2\\
15 & 25 & 1 & 3 & 26 & 1 & 3 & 24 & 1 & 2 & 23 & 1 & 2\\
20 & 29 & 1 & 4 & 28 & 1 & 3 & 26 & 1 & 4 & 25 & 1 & 3\\
30 & 32 & 1 & 3 & 32 & 1 & 4 & 29 & 1 & 4 & 30 & 1 & 3\\
50 & 61 & 3 & 3 & 55 & 2 & 4 & 43 & 2 & 4 & 43 & 2 & 5\\
\bottomrule
\end{tabular}
}
    \label{tab:ratio_eff_corr}
\end{table}

\subsection{Neutrino mass model}
\label{subsec:sgnmodel}

The signal yield for each heavy-neutrino mass hypothesis is determined from a binned maximum-likelihood fit to the invariant mass $m(\muonN\jet)$.
In the fits, the normalisation channel yield, the efficiency ratio, and background yields are Gaussian-constrained to their expected values within uncertainties.

The yields for the main background components are determined in the respective control regions.
The yields for $\W\to\mu\neu$ and $\Z\to\muon\muon$ background contributions are obtained from a binned maximum-likelihood fit of the invariant mass $m(\muonN\jet)$ in the \W region, and for the $\bbbar$ background in the $\bbbar$ region.
The fits in the control regions are performed separately for positively and negatively charged \muonW and per year of data taking with templates obtained from simulation.
The expected background yields in the signal region are determined by scaling the fitted yields according to simulation. 
The light QCD contribution in the signal region is estimated with a different method.
The efficiency of the \muonW uBDT requirement $\varepsilon_{\text{QCD}}$ is evaluated using the normalisation channel, assuming that it factorises from the other selection criteria that suppress the QCD background.
The number of light QCD events in the QCD region is obtained by subtracting from the total number of events the expected yields for the \W, \Z and heavy flavour background components.
The result is scaled by the ratio $\varepsilon_{\text{QCD}} / (1-\varepsilon_{\text{QCD}})$ to determine the number of light QCD events in the signal region.
The estimated background yields in the signal region are collected in \tab\ref{tab:crextrapolations} for same-sign and opposite-sign muons in Run 1 (2011 and 2012 combined) data. 
The uncertainty is dominated by the limited size of the simulated samples. 
The background predictions are tested in validation regions.
These are defined by inverting one by one the requirements of \tab\ref{tab:bkgcuts} defining the signal region.
The results are found to be in good agreement with the data.

\begin{table}[bt]
  \caption{Extrapolated background yields in the signal region for same-sign and opposite-sign muon channels. The uncertainty is statistical.}
    \label{tab:crextrapolations}
    \centerline{\begin{tabular}{l 
>{\collectcell\num}r<{\endcollectcell} @{${}\pm{}$} >{\collectcell\num}l<{\endcollectcell}
>{\collectcell\num}r<{\endcollectcell} @{${}\pm{}$} >{\collectcell\num}l<{\endcollectcell}
}
\toprule
Background &  \multicolumn{2}{c}{Same sign} & \multicolumn{2}{c}{Opposite sign}\\
\midrule
  \W\to\muon\neu   & 1.8 & 1.3 &    2.7 &   1.6\\
\bbbar             & 1.7 & 1.7 &    1.7 &   1.7\\
\Z\to\muon\muon    & 1.3 & 0.6 &   2251 &   161\\
light QCD          & 0.3 & 1.4 &    3.1 &   5.4\\
\bottomrule
\end{tabular}
}
\end{table}

The templates for both signal and background contributions are determined from simulation. 
The light QCD background is assumed to have the same shape as the $\bbbar$ background and therefore a single component is included in the fit for both.

\subsection{Results}
\label{subsec:results}
\begin{figure}[t!]
  \centering
  {\includegraphics[width=0.48\textwidth]{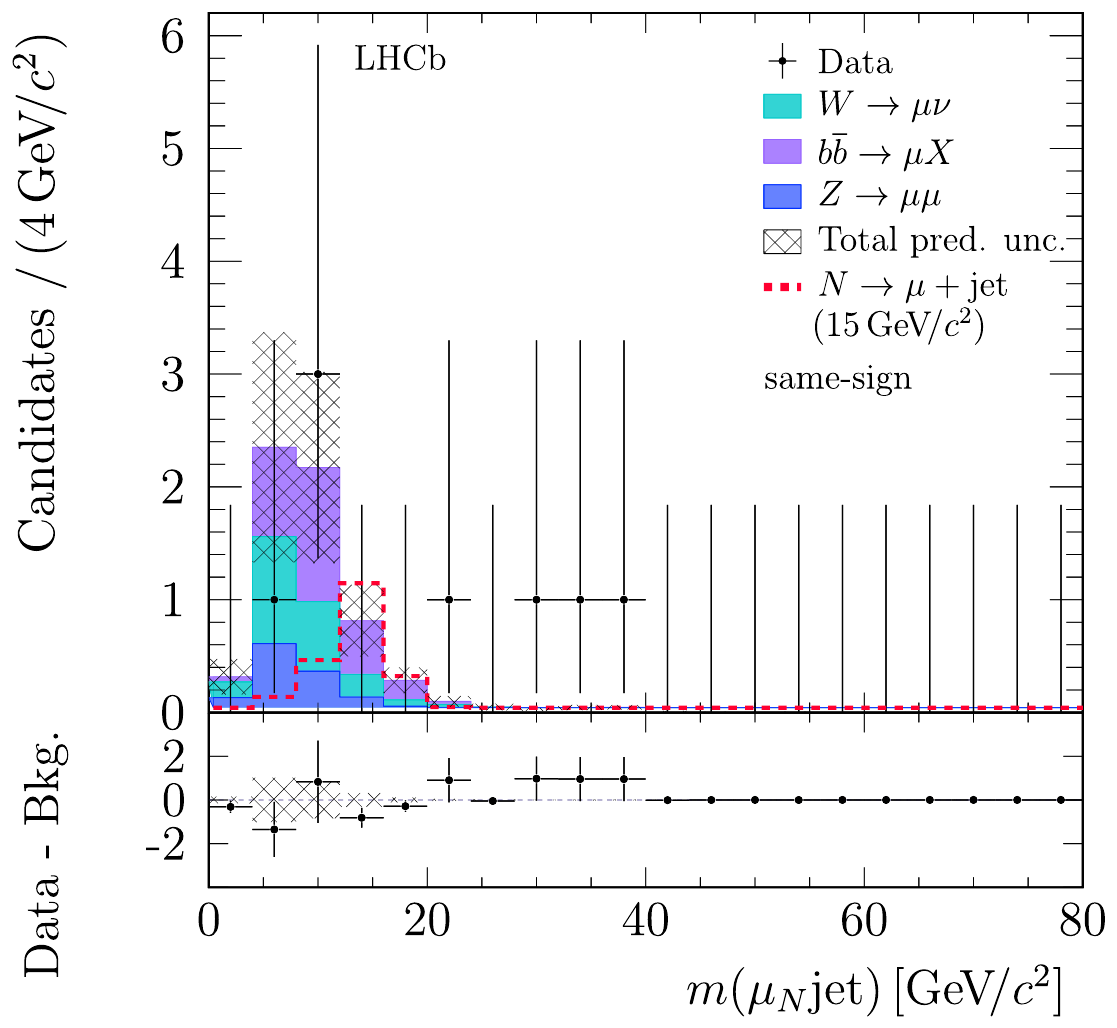}}\hspace{2mm}
  {\includegraphics[width=0.48\textwidth]{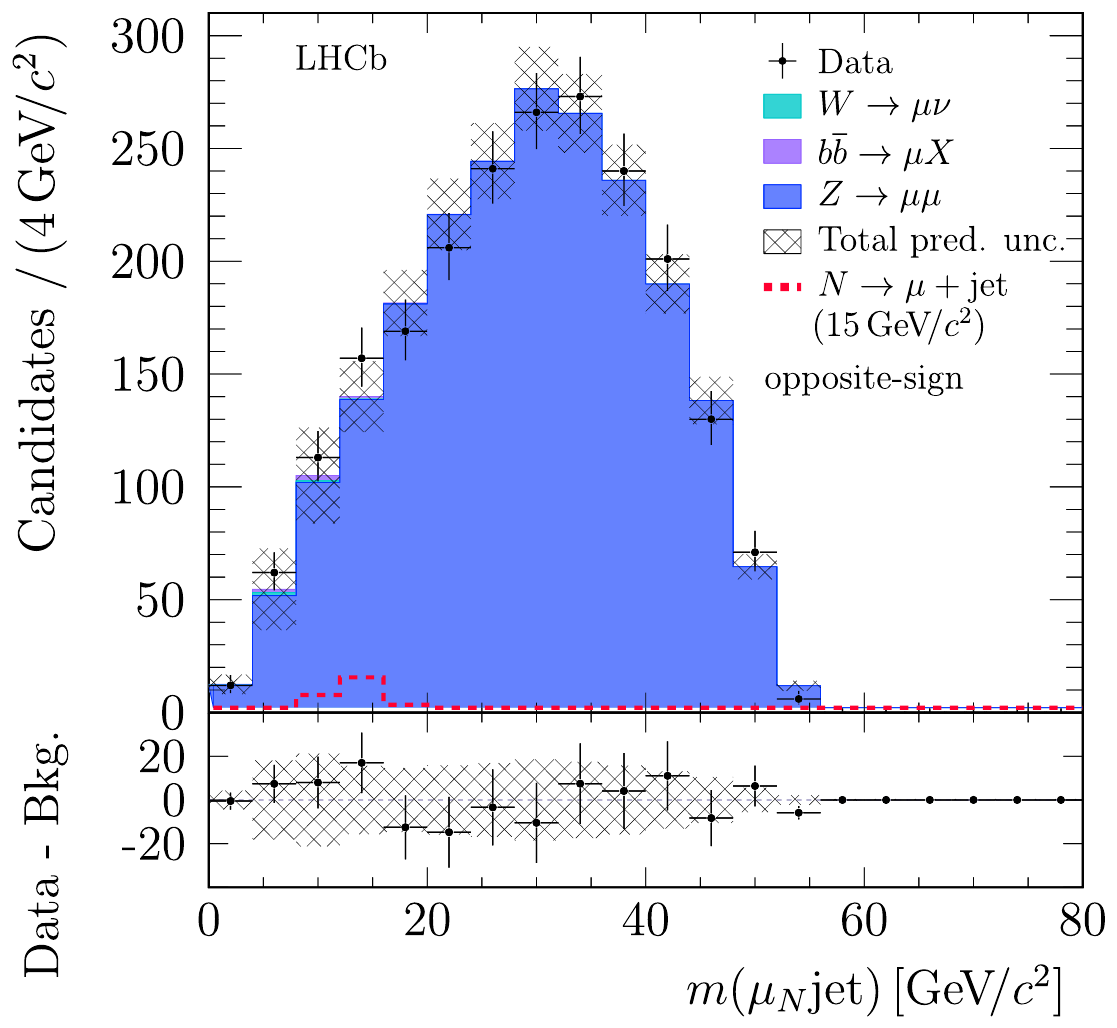}}
  \caption{Distributions of the invariant mass $m(\muon_N\jet)$ for (left) same-sign and (right) opposite-sign muons.
  The signal component corresponds to a $15\gevcc$ neutrino.
  }
  \label{fig:sr_fits}
\end{figure}

\begin{figure}[t!]
  \centering
  {\includegraphics[width=0.48\textwidth]{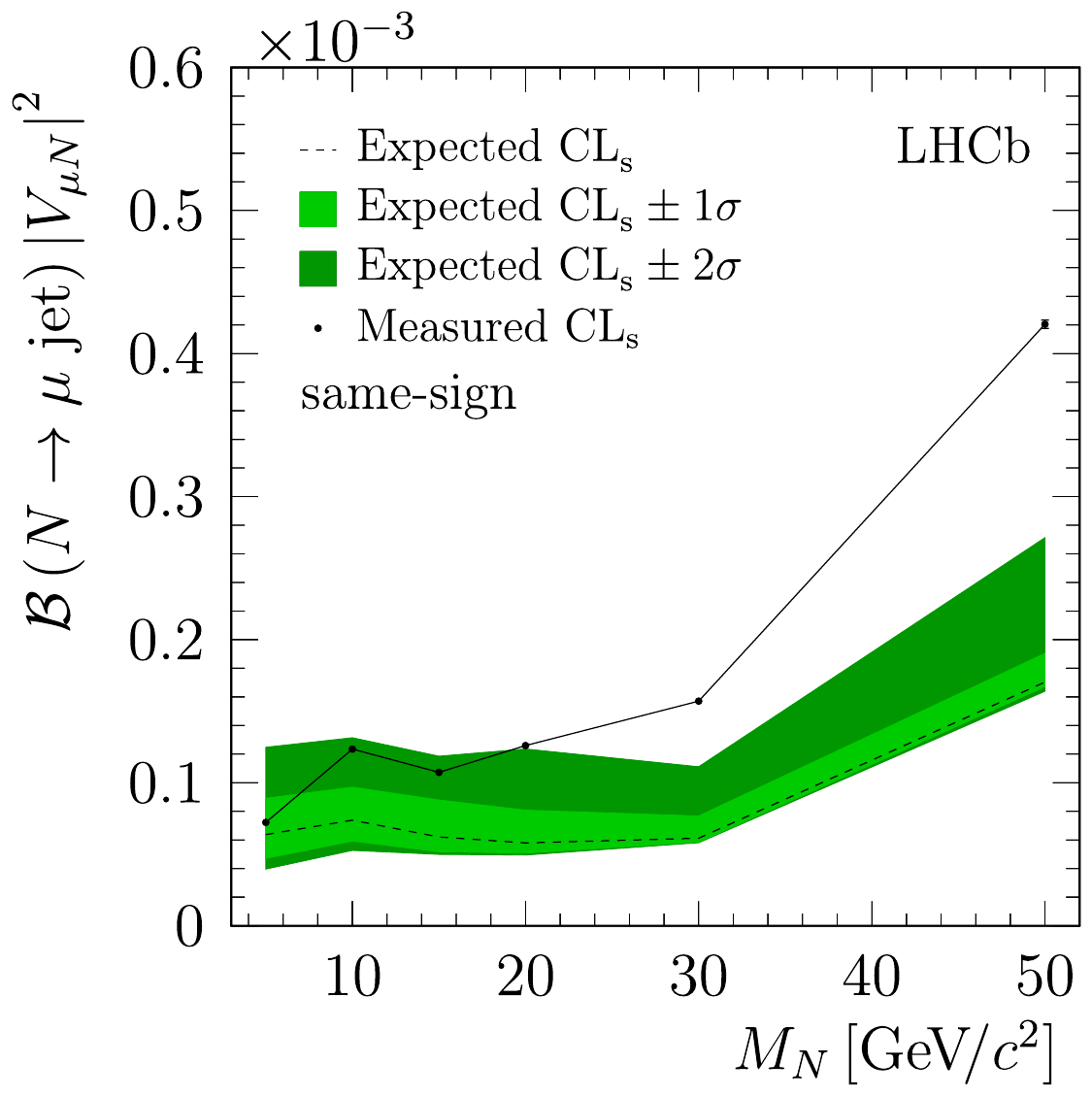}}\hspace{2mm}
  {\includegraphics[width=0.48\textwidth]{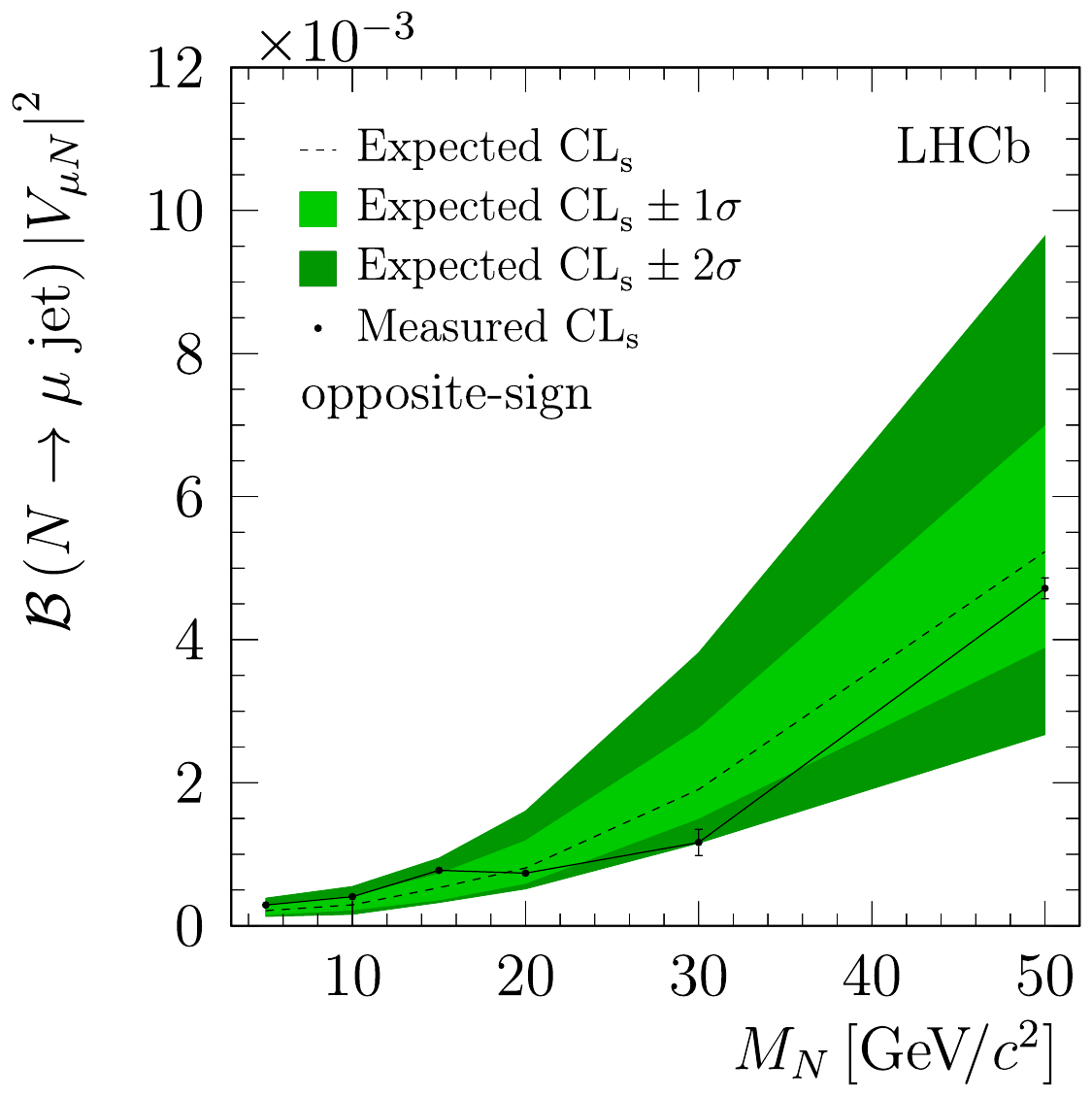}}
  \caption{Expected (dashed line) and observed (solid line) upper limit on $\BR(N\to\muon\jet)\left| V_{\muon    N}\right|^2$ at 95\% C.L. for (left) the same-sign muons sample and (right) the opposite-sign muons sample.
  The light and dark green bands show the $1\sigma$ and $2\sigma$ uncertainties, respectively, on the expected upper limits.
  }
  \label{fig:unblinded_ulvsmass}
\end{figure}

The number of events observed in data in the signal region amounts to 8 and 2147 for same-sign and opposite-sign muons, respectively.
A single fit to the Run 1 data is performed since the 2011 and 2012 templates are found to be compatible.
The distributions of the invariant mass $m(\muon_N\jet)$ for same-sign and opposite-sign muon data are shown in \fig\ref{fig:sr_fits} with the fits for the $15\gevcc$ neutrino mass hypothesis superimposed.
Upper limits at 95\% confidence level on $\BR(N\to \muon \jet) \left| V_{\muon N}\right|^2$ are set for each heavy-neutrino mass hypothesis using the CLs method~\cite{Read:2002hq} with a one-sided profile likelihood ratio~\cite{Cowan:2010js} as test statistic. 
The upper limits as a function of heavy-neutrino mass are shown in \fig\ref{fig:unblinded_ulvsmass}.
For the same-sign muons sample and neutrino mass in excess of $20\gevcc$, the measured limit is between 2 and 3.8 standard deviations above the expected limit.
The worse limit obtained with respect to the expectation can be attributed to the four data candidates with $m(\muon\jet)$ between $20$ and $40\gevcc$.
The value of the muon identification BDTs for three of the candidates are very close to the requirements, defined a priori with a blinded procedure, indicating that they are background-like and probably a QCD fluctuation.
Each candidate has also a relatively large value for the missing transverse momentum in the event, which is not characteristic for the signal. 
Consequently, the excess at high mass is likely the result of an imperfectly modelled component of the background.
For the opposite-sign muons samples, the expected limit is a factor 5 to 10 worse due to the irreducible background from Drell-Yan processes, in agreement with expectations.

\begin{figure}[t!]
  \centerline{\includegraphics[width=0.9\textwidth]{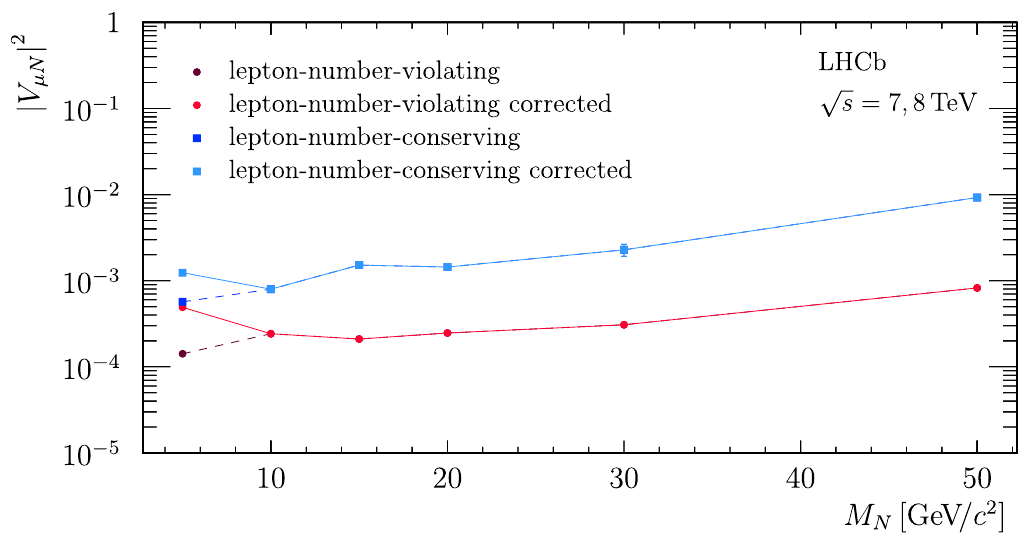}}
  \caption{Observed upper limit on the mixing parameter $\left| V_{\muon    N}\right|^2$ between a heavy neutrino and a muon neutrino in the mass range $5-50 \gevcc$ for same-sign and opposite-sign muons in the final  states with and without lifetime correction.
  }
  \label{fig:sum_limits}
\end{figure}

To set upper limits on the coupling, the results of \fig\ref{fig:unblinded_ulvsmass} are scaled by \mbox{$\BR(N\to\muon\jet)=0.51$}, computed as described in \sect\ref{sec:introduction} assuming $\left| V_{\electron N}\right|^2 = \left| V_{\tauon N}\right|^2 = 0$.
For the $5\gevcc$ heavy-neutrino mass hypothesis, at the limit set, the heavy neutrino is expected to be long-lived with a lifetime of $3.8\ps$ and $1.1\ps$ for same- and opposite-sign muons in the final states, respectively. 
Since this search targets prompt heavy neutrinos, the acceptance is corrected accordingly.
The constraints on the coupling as a function of mass for the opposite- and same-sign muons final state, with and without the acceptance correction factor applied, are illustrated in \fig\ref{fig:sum_limits}.

\section{Conclusion}
\label{sec:conclusion}

A search for a prompt heavy neutrino in the decay $N\to\mu\jet$ is performed using data from proton-proton collisions recorded by the LHCb experiment, corresponding to a total integrated luminosity of $3\invfb$.
No evidence for heavy neutrinos is observed  
and limits of the order of $10^{-4}$ and $10^{-3}$ are set as a function of heavy-neutrino mass for lepton-number-conserving and lepton-number-violating decays, respectively.
An upwards fluctuation is present in the lepton-number-violating case, which is likely ascribable to an imperfectly modelled component of the background.
These represent the first limits on the coupling to a heavy neutrino in the mass range 5-$50\gevcc$ at LHCb.
For the first time the signature of two muons and a low mass jet has been probed for heavy neutrinos with mass lower than $20\gevcc$.
Furthermore, this is the first limit on lepton-number-conserving decays of a prompt heavy neutrino in the mass range of interest.
The observed limits on lepton-number-violating decays are not yet competitive with the existing limits~\cite{Abreu:1996pa,Sirunyan:2018mtv,Aad:2019kiz}. %
With an integrated luminosity of $50\invfb$, a better sensitivity than the current most stringent limit could be reached for the same-sign muons channel.
While this analysis targets prompt heavy-neutrino decays, better sensitivity for low heavy-neutrino masses can be achieved by including long-lived signatures.

%
\section*{Acknowledgements}
%
%
\noindent We would like to thank Dr. Brian Shuve of the Harvey Mudd College for the help with the event generation model and for cross-checking the calculations shown in \fig\ref{fig:HNBR}.
We express our gratitude to our colleagues in the CERN
accelerator departments for the excellent performance of the LHC. We
thank the technical and administrative staff at the LHCb
institutes.
We acknowledge support from CERN and from the national agencies:
CAPES, CNPq, FAPERJ and FINEP (Brazil); 
MOST and NSFC (China); 
CNRS/IN2P3 (France); 
BMBF, DFG and MPG (Germany); 
INFN (Italy); 
NWO (Netherlands); 
MNiSW and NCN (Poland); 
MEN/IFA (Romania); 
MSHE (Russia); 
MICINN (Spain); 
SNSF and SER (Switzerland); 
NASU (Ukraine); 
STFC (United Kingdom); 
DOE NP and NSF (USA).
We acknowledge the computing resources that are provided by CERN, IN2P3
(France), KIT and DESY (Germany), INFN (Italy), SURF (Netherlands),
PIC (Spain), GridPP (United Kingdom), RRCKI and Yandex
LLC (Russia), CSCS (Switzerland), IFIN-HH (Romania), CBPF (Brazil),
PL-GRID (Poland) and OSC (USA).
We are indebted to the communities behind the multiple open-source
software packages on which we depend.
Individual groups or members have received support from
AvH Foundation (Germany);
EPLANET, Marie Sk\l{}odowska-Curie Actions and ERC (European Union);
A*MIDEX, ANR, Labex P2IO and OCEVU, and R\'{e}gion Auvergne-Rh\^{o}ne-Alpes (France);
Key Research Program of Frontier Sciences of CAS, CAS PIFI,
Thousand Talents Program, and Sci. \& Tech. Program of Guangzhou (China);
RFBR, RSF and Yandex LLC (Russia);
GVA, XuntaGal and GENCAT (Spain);
the Royal Society
and the Leverhulme Trust (United Kingdom).

\addcontentsline{toc}{section}{References}
\bibliographystyle{LHCb}
\bibliography{main,standard,LHCb-PAPER,LHCb-CONF,LHCb-DP,LHCb-TDR}

\newpage
\centerline
{\large\bf LHCb collaboration}
\begin
{flushleft}
\small
R.~Aaij$^{31}$,
C.~Abell{\'a}n~Beteta$^{49}$,
T.~Ackernley$^{59}$,
B.~Adeva$^{45}$,
M.~Adinolfi$^{53}$,
H.~Afsharnia$^{9}$,
C.A.~Aidala$^{84}$,
S.~Aiola$^{25}$,
Z.~Ajaltouni$^{9}$,
S.~Akar$^{64}$
% ...

{\footnotesize \it

% ...
$ ^{9}$Universit{\'e} Clermont Auvergne, CNRS/IN2P3, LPC, Clermont-Ferrand, France\\
% ...
$ ^{25}$INFN Sezione di Milano, Milano, Italy\\
% ...
$ ^{31}$Nikhef National Institute for Subatomic Physics, Amsterdam, Netherlands\\
% ...
$ ^{45}$Instituto Galego de F{\'\i}sica de Altas Enerx{\'\i}as (IGFAE), Universidade de Santiago de Compostela, Santiago de Compostela, Spain\\
% ...
$ ^{49}$Physik-Institut, Universit{\"a}t Z{\"u}rich, Z{\"u}rich, Switzerland\\
% ...
$ ^{53}$H.H. Wills Physics Laboratory, University of Bristol, Bristol, United Kingdom\\
% ...
$ ^{59}$Oliver Lodge Laboratory, University of Liverpool, Liverpool, United Kingdom\\
% ...
$ ^{64}$University of Cincinnati, Cincinnati, OH, United States\\
% ...
$ ^{67}$Syracuse University, Syracuse, NY, United States\\
% ...
$ ^{84}$University of Michigan, Ann Arbor, United States, associated to $^{67}$\\
\bigskip
% ...
\medskip
}
\end{flushleft}

\end{document}